# Performance studies of a pixel tracker in the Silicon Detector (SiD) concept for a future linear collider

**T White**

Particle Physics Department, Rutherford Appleton Laboratory,

E-mail: gw236@bath.ac.uk

**Abstract.** Using the simulation framework of the SiD detector to study the $H \rightarrow \mu^+\mu^-$ decay channel showed a considerable gain in signal significance could be achieved through an increase in charged particle momentum resolution. However more detailed simulations of the $Z \rightarrow \mu^+\mu^-$ decay channel demonstrated that significant improvement in the resolution could not be achieved through an increase in tracker granularity. Conversely detector stability studies into missing/dead vertex layers using longer lived particles displayed an increase in track resolution. The existing 9.15 cm x 25 μm silicon strip geometry was replaced with 100 μm x 100 μm silicon pixels improving secondary vertex resolution by a factor of 100. Study into highly collimated events through the use of dense jets showed that momentum resolution can be increased by a factor of 2, greatly improving signal significance but requiring a reduction in pixel size to 25 μm. An upgrade of the tracker granularity from the 9.15 cm strips to micrometer sized pixels requires an increase in number and complexity of sensor channels yet provides only a small improvement in the majority of linear collider physics.

## 1. Introduction and theory

*1.1. Introduction to particle physics*
Particle physics is the study of elementary particles and their interactions [1]. The best theoretical description of particle physics is the Standard Model (SM) and it is extremely successful at describing the fundamental constituents of matter as well as three of the four known interactions. It consists of 12 fermions (6 quarks and 6 leptons) which make up all matter and 4 vector bosons which through exchange between fermions mediate the electromagnetic, strong and weak interactions. Figure 1 shows a summary of the fundamental constituents of the SM. Gravity has yet to be included in the SM framework.

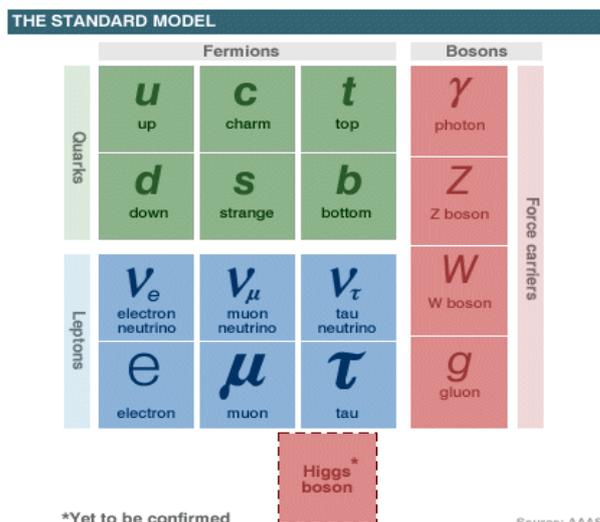

Figure 1. Overview of the fundamental constituents of the Standard Model [1].

Performance studies of a pixel tracker in the Silicon Detector (SiD) concept for a future linear collider

One of the building blocks of the SM is the unification of the electromagnetic and weak interactions. It is currently understood that the separation of these interactions at lower energies is due to the different masses of the force carriers (bosons). The electromagnetic interaction is carried by the massless photons and the weak interaction by the massive W and Z bosons [2]. In the SM this lack of symmetry between the interactions is resolved by the Higgs mechanism which leads to the prediction of a new scalar boson with coupling proportional to mass. The Higgs boson is the only fundamental particle of the SM which has not yet been experimentally verified and a crucial challenge in particle physics is to explore large energy scales to discover the Higgs and in doing so expose the nature of electroweak symmetry breaking [3]. The SM features some limitations, most obviously the exclusion of gravitational interactions. Another assumption of the SM is that neutrinos are assumed to be massless; however there is strong evidence that neutrinos have finite mass. Additionally the SM also contains 19 free parameters, of which little is known of the origin and necessity. Attempts to understand major features of the early universe such as dark matter and matter-antimatter asymmetry will require new physics beyond that of the SM [4]. These are questions that experimentalists hope to resolve in the near future through the study of high energy particle collisions, this achieved through the use of large purpose built particle colliders and detectors [5].

The Large Hadron Collider (LHC) had its first collisions on the 23$^{rd}$ November 2009 and shortly after set a new world record with two beams at 1.05 TeV [6]. If the properties of the Higgs fit with SM predictions its existence is expected to be confirmed within three years [7]. Due to the timescales involved in the development of a particle collider, typically one or two decades, research has already begun into the next step in high energy physics. One possible future development is the construction of an electron-positron linear collider specifically designed to precisely study the physics discovered at the LHC [8]. This project looks at the design of one such detector in development for a future linear collider and benchmarks the performance improvement that can be achieved with a finer segmented tracking system. It will show how greater precision can be achieved in the measurement of the trajectory and momentum of charged particles.

*1.2. The need for precision science and a lepton linear collider*
The engineering behind particle colliders has driven the development of particle physics over the last 40 years. Recent experiments at colliders have discovered the W and Z bosons (UA1 + UA2, SPS CERN), and most recently the top quark (Tevatron, Fermilab) [2]. This required collaboration between high energy hadron colliders to search for unknown physics at higher energy levels and more precise finely tuned lepton colliders to probe into the properties of any newly discovered particles.

**Table 1.** Higgs branching ratios for Higgs Mass = 120 GeV.

| Decay Products | Standard Model | MSSM case 1 | MSSM case 2 |
| --- | --- | --- | --- |
| bb | 68% | 74% | 70% |
| WW | 13% | 10% | 12% |
| gg | 7% | 5% | 7% |
| $\tau\tau$ | 7% | 8% | 7% |
| cc | 3% | 2% | 3% |
| ZZ | 2% | 1% | 1% |
| other | <1% | <1% | <1% |

Over the next couple of years the LHC will give us a first look at TeV scale physics and this is expected to include the discovery of the Higgs [7]. However finding the Higgs is only half the challenge as measuring the properties accurately is necessary to help distinguish between different theoretical models. This consists of high precision measurements of the Higgs mass, width, spin, and branching ratios. The precision required in these measurements must be sufficient enough to discriminate between any competing theories. As can be seen from table 1 the branching ratios can be very different between theoretical models. However at the LHC branching ratios will be measured with 10% precision at best, which will not allow for sufficient discrimination between different theoretical models [9]. This is known in as the LHC inverse problem. In addition to this many of the rarer Higgs decay channels are not accessible at all in the LHC due to overwhelming backgrounds (See 1.4).

A lepton collider has several advantages over a hadron collider. Firstly it has a well defined interaction point (IP) of the order of a few micrometers which is many orders of magnitude smaller than the LHC currently is able to obtain. A lepton collider also has a well defined centre of mass (CMS) energy. This is due to the collision between two point-like particles, whereas in a hadron collider the collision is between two composite particles. This cleaner environment aids considerably during event reconstruction. The reason for a

Performance studies of a pixel tracker in the Silicon Detector (SiD) concept for a future linear collider linear collider follows from the decision to collide leptons. Synchrotron radiation occurs when relativistic charged particles move through magnetic fields. At high energies leptons are highly relativistic due to their small mass and the emitted photons represent considerable energy loss through synchrotron radiation. This loss scales according to equation (1), where ΔE is the energy radiated per particle per turn, $\rho$ is the bending radius, $\beta$ is the particle velocity and $\gamma = (1 - \beta^2)^{-1/2}$ [4]. An example of this is the large electron positron (LEP) collider which operating at CMS energy of 91.2 GeV radiated 1.2 MW whereas at 200 GeV it has already increased to around 30 MW [10]. Even without acceleration this amount of energy must be supplied to maintain the energy of the leptons. Additionally at large values of CMS energy this can become a problem through damage to electronics and infrastructure [10].

$$\Delta E = \frac{4\pi}{3}\left(\frac{e^2 \beta^3 \gamma^4}{\rho}\right) \quad (1)$$

Two electron/positron linear collider concepts are currently in development, the International Linear Collider [8] (ILC) and the Compact Linear Collider [11] (CLIC). The Silicon Detector concept [12] (SiD) that this project will focus on is a detector that has been designed to make precision measurements while maintaining sensitivity to a wide range of possible new phenomena that may be present at a future linear collider. SiD is being developed with ILC in mind but the design and physics processes behind this concept are in general applicable to both.

*1.3. Detector physics, infrastructure and design*
SiD is a general-multipurpose detector with $4\pi$ solid angle coverage to measure the energy and momentum of particles with high precision. It consists of multiple task-orientated detectors, see figure 2(a). At the very centre, closest to the IP is the vertex detector which consists of a barrel of 5 layers as well as a forward and backward disk region each consisting of 4 layers all tiled with 20 μm silicon pixels (finely segmented regular quadrilateral sensors) and giving a very high spatial resolution. The tracker, located around the vertex detector, also consists of a barrel with 5 layers and 4 disks on each end (known as the endcaps) and is tiled with silicon strips 25 μm by 9.15 cm long. In addition to this 3 silicon pixel outer disks provide coverage in the transition region between the vertex detector and the tracker. The entire vertex detector and tracker can be seen in figure 2(b). This set up allows charged particles to be reconstructed very accurately by reconstructing tracks through hits left in the vertex detector and tracker.

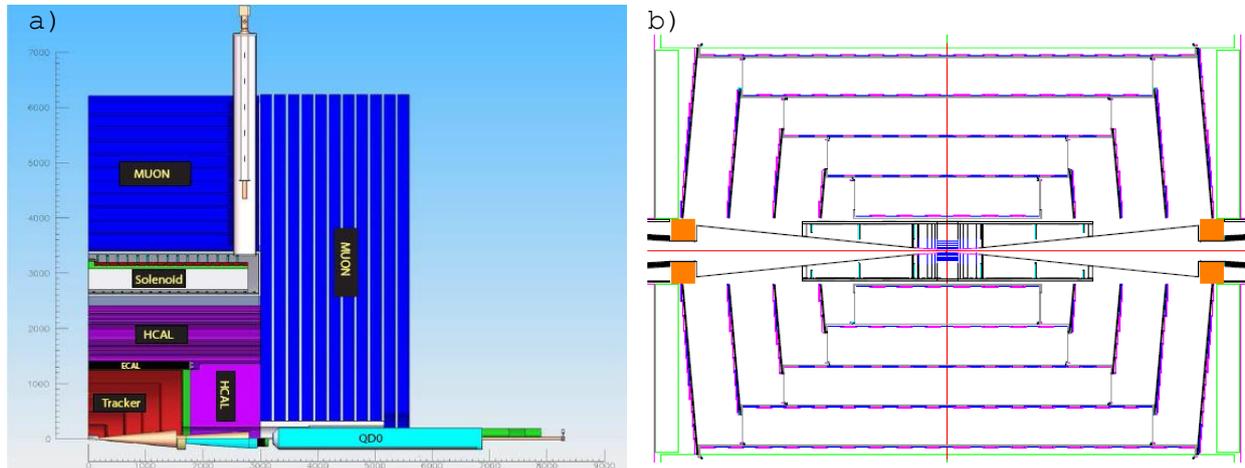

Figure 2. (a) Overview of the SiD detector geometry and (b) R-z view of the vertex detector and tracker arrangement [12].

Situated outside of the tracker are the electromagnetic (ECAL) and hadronic (HCAL) calorimeters which measure the energy deposition of the particles. These will consist of a higher granularity than conventional calorimeters and be placed inside the surrounding solenoid. This allows for calorimeter aided tracking and the use of Particle Flow Algorithms [13] which are beyond the scope of this project but crucial to the overall design of SiD. In this project the calorimeters will be presumed to supply only energy readings for the reconstructed particles. A large 5 T superconducting solenoid surrounds the tracker and calorimeter providing the necessary magnetic field to curve the charged particle tracks allowing momentum measurements and charge identification. Muons, long lived particles and neutrinos have a low enough cross section to penetrate the tracker, calorimeters, solenoid and the necessary services. Hence the rest of the detector consists of the muon detectors, large resistive plate chambers which enclose the entire detector to identify the muons and long lived particles. Finally the presence of neutrinos which escape the detector without interaction is usually inferred from an imbalance of momentum/energy [2]. This project is mainly focused on the modification of the silicon tracker. The current design is built around silicon strips 9.15 cm by 25 μm. The aim of this project is to benchmark the performance increase obtained by using micrometer sized pixels across the entirety of the tracker in a similar way to the vertex detector.

*1.4. Higgs detection and benchmarking channels*
A few decay channels are used to benchmark the performance of the proposed tracker. The $H \rightarrow \mu^+\mu^-$ channel is used to study the rare decay of a Higgs produced via the Higgs-Strahlung process into two muons and the Z into neutrinos ($e^+e^- \rightarrow H^0Z^0 \rightarrow \mu^+\mu^-\nu\bar{\nu}$). This rare channel has a branching ratio of the order of 0.01% but due to the precision available in the detector it should be possible to filter out the background to obtain a statistically significant sample of signal events. This channel would allow for high precision measurement of the intrinsic Higgs properties. The similar and well documented $Z^0 \rightarrow \mu^+\mu^-$ decay channel is also used to study simple charged particle track reconstruction from hits made in the vertex detector and tracker. Long lived particles decay away from the IP producing secondary vertices, these must be rebuilt using limited layers from the vertex detector and often using purely tracker information. These events are analysed simulating both the possibility of malfunctioning layers (stability studies) as well as long lived particles themselves. $K_S^0$ particles with a lifetime of around $10^{-10}$ s travel tens of centimetres (missing vertex layers) before decaying into two charged pions. This allows for a direct evaluation of the performance of the tracker in the reconstruction of charged particle tracks. Finally dense jets are used to evaluate the reconstruction of more highly collimated events. These high track density events make reconstruction more difficult due to multiple hits in single sensors. This leads to the loss of hits which can cause the loss of tracks during reconstruction and hence a loss in total momentum. This can lead to incorrectly reconstructed neutrinos.

*1.5. Previous studies*
A similar study has previously attempted to ascertain the necessary momentum resolution required for Higgs discovery and adequate measurement of its properties. H.J.Yang carried out a study on the detector design in 2007 [14] using a simplified computational method known as "fast MC" (See 2.4). Looking purely at the rare $e^+e^- \rightarrow H^0Z^0 \rightarrow \mu^+\mu^-X$ decay the conclusion was that the "SiD tracker with nominal track momentum resolution makes it possible but still hard to measure … but direct measurement is feasible if the track momentum resolution is improved by a factor of 2". This study will take this investigation further and extend it to new Higgs decay modes.

**2. Computational methods**
During the development of any detector the entire process is simulated. This begins with Monte Carlo (MC) event generation which produces descriptions of the particles created during a lepton collision. The descriptions of these particles are passed through a detector simulation reproducing the interaction of the particles with matter. Charge deposition and ionization is simulated along with the read out electronics in a process known as digitization. This produces 'hits' in the simulated detector layers which can be used to reconstruct the original event. Event reconstruction involves fitting helical tracks to the 'hits' and identifying the properties of the charged particles. Finally analysis programs are used that compare the reconstructed events to the original MC events to evaluate the success of the detector and reconstruction. The data produced is used to influence the detector design.

*2.1. Monte Carlo event generation*
Data samples consisting of a large number of particle collision events are generated by the WHIZARD Monte Carlo program [15]. WHIZARD was chosen as it can implement many physical models, including the SM and due to the detailed description of beam properties such as energy spread, crossing angle and beamstrahlung it is particularly adapted to linear collider physics. The physics behind WHIZARD is given by the SM theory behind strong and electroweak interactions and has been shown to reliably compute cross sections for 4, 6 and 8 fermion events. A second program PYTHIA [16] is then used for final state quantum electrodynamic (QED) and quantum chromodynamic (QCD) calculations, including parton (quarks and gluons) showering, fragmentation and decay to final state observable particles. PYTHIA contains a built in library of many leading order processes from analytical results as well as various phenomenological models which are used in the simulation. The simulated events are known as the "MC truth" events.

Performance studies of a pixel tracker in the Silicon Detector (SiD) concept for a future linear collider

## 2.2. Detector simulation

Simulation of the detectors response to the generated final state events is handled by the GEANT4 toolkit [17]. GEANT4 is an object orientated package designed to simulate the passage of particles through matter. It can handle all aspects of the simulation from the geometry of the system, the materials involved, the tracking of particles through materials and external magnetic fields, the physics processes governing particle interactions and the response of sensitive detector components [18].

The geometry used in GEANT4 is input via an XML file at runtime and represents a simple geometric description of the location of the layers of the detector. These layers are defined as "sensitive elements" of the detector and as charged particles pass through these layers ionization and charge deposition is simulated. If the particle energy is above a certain threshold then the location of that particle is recorded as a 'hit'. The hits returned from the detector simulation are located at the exact position at which the particle passed through and ionized that particular layer. In reality the layers of the detector will be tiled with small segments known as pixels or strips. The electronics associated with this will read out any reported hit within that segment (digitization). In simulation the process of tiling the detector with sensors is known as hit segmentation and is the basis for this project. Segmentation transforms each "hit" from the detector simulation, and using the position and size of the sensors reads out a new hit now positioned at the centre of the sensor, this gives the detector an intrinsic hit resolution.

## 2.3. Track and particle reconstruction

The trajectory of a charged particle travelling through a magnetic field can uniquely be defined by a helix. The fitting of a helix through the hits left in the tracker layers is known as track reconstruction. A track finding strategy contains all possible combinations of layers that could have been made by a single charged particle; the strategy is then used to reconstruct tracks from the detector hits. The layer combinations are stored in an XML file that can be manually edited to give the user a high level of precision over track reconstruction; however in reality strategy builders produce these files after analysis of the detector configuration and a selection of MC events. The track reconstruction algorithm currently bases all decisions on a global chi-squared ($\chi^2$) measure of success [19]. The $\chi^2$ fit of the helix to the contributing hits is calculated and compared to a maximum value set by the track finding strategy. A fast helix fitter takes the tracker hits and produces a helix. This is done by fitting a circle to the x and y coordinates of the hits and three parameters $\omega$, $\phi_0$ and $d_0$ are determined. A line fit is then performed along a plane in the z direction to determine $z_0$ and $\tan(\lambda)$. These five parameters uniquely define a helix according to figure 3.

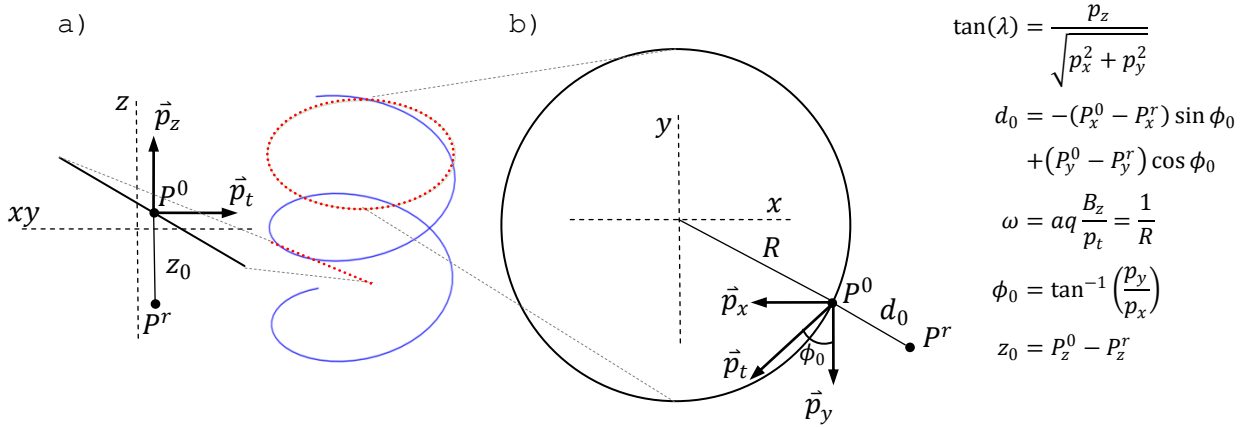

Figure 3. Parameterization of a helix into a line in the z-direction (a) and a circle in the x-y plane (b) along with 5 parameters necessary to uniquely identify the helix.

Tracker hits supply 3 spatial values and as 5 parameters are required to uniquely define the helix it is necessary to have at least 2 hits. However three types of hit are supported by the track reconstruction algorithm corresponding to the three types of sensors that could be tiled onto the detector layers. Firstly, micrometer pixel (square) hits that have two measured co-ordinates, secondly, strip hits which have one measured and one bounded co-ordinate and thirdly stereo hits formed from a pair of crossed strip hits. These stereo hits are produced from two strips in consecutive layers where the second layer is offset from the first by the stereo angle. In all cases the third co-ordinate is taken from the layer of the detector in which the sensor lies. Reconstruction from 2 hits is only possible with pixel hits or stereo pair hits each of which supply 2 coordinates and the third can be indirectly obtained from the layer. A strip hit however supplies only one coordinate directly (short axis) and one that can be indirectly determined from the layer leaving the third coordinate only bounded by the long axis of the strip. In this case the bounds of multiple strip hits consecutively placed in the detector can be overlapped giving an allowed polygonal region the centre of which is used to compute the global $\chi^2$.

## 2.4. The "fast MC" approach

Due to the CPU time required to simulate the passage of every particle through the detector another less robust and less accurate technique is sometimes used. This method known as "fast MC" parameterizes the momentum resolution of the charged "MC truth" particles according to equation 2. The two terms in the equation are the energy loss term and the multiple scattering term respectively. The energy loss term represents energy loss through ionization and charge deposition while the multiple scattering term represents the effect from the many scattering centres in the detector layers. The variables $a$ and $b$ represent the relative strength of each term and both depend on the detector material. This method produces a particle track from the charged "MC truth" particles taken directly from PYTHIA without the need for the full detector simulation (2.2) or track reconstruction (2.3) [14].

$$\left(\frac{\Delta P_t}{P_t}\right)^2 = a^2 + \left(\frac{b}{P_t \sin\theta}\right)^2 \qquad (2)$$

## 2.5. Benchmarking and analysis

The invariant mass is an important physical quantity which remains invariant under boost, and is hence the same in all reference frames. It is calculated using the norm (in Minkowski space) of the particle four vector containing energy and momentum and for each particle this produces a peak in the mass spectrum [4]. The width of the invariant mass peak may be thought of as an accurate indication of the resolution of the detector. In the $H \to \mu^+\mu^-$ decay channel the combination of the four vectors from the two reconstructed decay products (the muons) can give the Higgs invariant mass. The invariant mass, known as signal, is then combined with SM background data in datasets that represent real sizes and timescales. Using the information in the events to perform cuts and classify the new dataset the background can be reduced whilst retaining as much signal as possible. A plot of the invariant mass of the remaining dataset will then produce a background shape (normally approximated to an exponential or flat polynomial within a local region around the Higgs mass, $m_H$) with a small peak at $m_H$ due to the signal events. A probability density function made up of a Gaussian and background shape can then be fitted to the data using the software library ROOT (RooFit) [20]. An unbinned maximum likelihood fit [21] gives the statistical significance of the signal shape. This is often known as sigma and used throughout particle physics to represent the significance of a discovery. This gives an idea of the improvement in the physics results that can be obtained with increased momentum resolution. The $H \to \mu^+\mu^-$ channel is smeared using the "fast MC" method.

The majority of the analysis was achieved using programs written in Java. These programs, known as drivers, build upon the large existing framework org.lcsim [22] which allows access to the reconstructed event files. JAS3 [23] is then used to run these drivers over a large number of events producing histograms representing direct or derived variables. The reconstructed files retain the "MC truth" event data produced by PYTHIA for comparison to the reconstructed particles. The development of a truth matching strategy allows a direct comparison between reconstructed tracks and MC charged particles by matching the two. This is achieved by analysing the MC particles responsible for the hits contributing to a particular track. Selecting tracks reconstructed from hits that come principally from a single charged MC particle is sufficient for simple truth matching. (See figure 4).

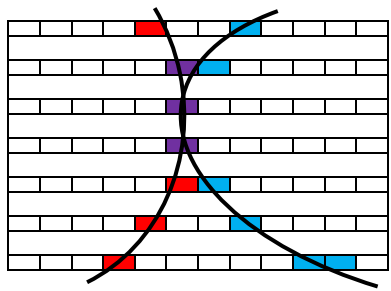

Figure 4. Hits from two different MC particles (red/blue) and two reconstructed tracks (lines). Both tracks can be correctly matched to the particle regardless of which particle/track the purple hits are associated.

For the full simulation the segmentation process is checked to ensure it has successfully altered the intrinsic spatial resolution of the detector. The positions of the individual hits will be compared before and after reconstruction in each Cartesian direction. These hit distributions will be the first check that the segmentation code runs successfully. Using a truth matching algorithm it will then be possible to analyse individual reconstructed tracks. Resolution histograms will be produced showing the change in momentum, angle or any of the track parameters mentioned in section 2.3. For the longer lived particles the variable used to judge success will be the position of the secondary vertex in the reconstructed events compared to the "MC truth". This variable is highly sensitive to small changes in hit position as it requires extrapolation back along the particle track towards the IP. In the reconstruction this is the point of closest approach on the track to the "MC truth" decay point. Secondary vertices have not been studied before so a new track finding strategy that is able to seed tracks using hits in the tracker was built and developed. It should also be noted that the majority of the computational work is achieved using a batch farm of multiple computers working in parallel.

## 3. Results and discussion

### 3.1. $H \rightarrow \mu^+\mu^-$ decay channel using "fast MC"

The first channel studied is the $H \rightarrow \mu^+\mu^-$ channel using the "fast MC" technique. Figure 5 shows the invariant mass reconstructed from the properties of the two final state muons. This shows that by choosing the parameters of equation (2) correctly the root mean squared (RMS) value of the invariant mass can be changed. For this initial study it is presumed that doubling the RMS of this value is equivalent to having a detector with twice the resolution. Figure 6(a) shows a sample of the Higgs invariant mass after being combined with background. The exponential background occurs from the $Z^0$ invariant mass peak which occurs at 91 GeV (to the left of figure). Using cuts and a properly trained classifier it is possible to reduce the number of background events, however it was found that the more significantly the background was reduced the more the background shape would be altered. In the most extreme cases it was found the background shape took on a Gaussian appearance with a peak at the Higgs mass. This situation makes it considerably more difficult to separate signal and background. Less drastic reduction in the background is often used leaving a locally flat shape either side of the Higgs mass peak, see figure 6(b).

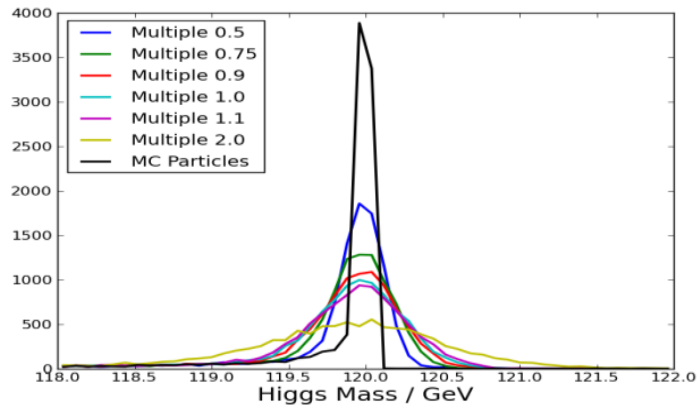

Figure 5. Higgs invariant mass after "fast MC" smearing of tracks calibrated with multiples of the current SiD Higgs invariant mass RMS.

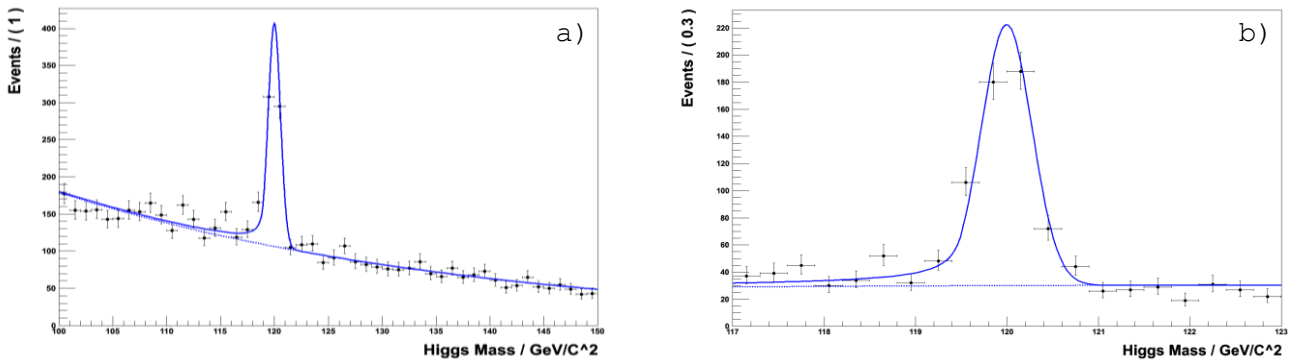

Figure 6. Fits from RooFit of a crystal ball [24] probability density function with an (a) exponential background and (b) locally flat background.

Performance studies of a pixel tracker in the Silicon Detector (SiD) concept for a future linear collider

Applying an unbinned maximum likelihood fit to the signal-background plots it is possible to get a statistical significance of the Higgs mass peak. This has been plotted in figure 7 for various detector resolutions and for differently sized datasets representing different detector running times. The confidence with which a signal can be said to be found depends strongly on the shape of the signal peak. The alteration of the momentum resolution shows considerable change in the signal significance achieved at different detector resolutions. In many cases it is possible to run for a shorter period and obtain a similar level of confidence in the results. As with the H.J.Yang study it seems upgrading the momentum resolution to 0.5 that of the current SiD design leads to significantly improved physics results and justifies a more detailed investigation into improving detector resolution.

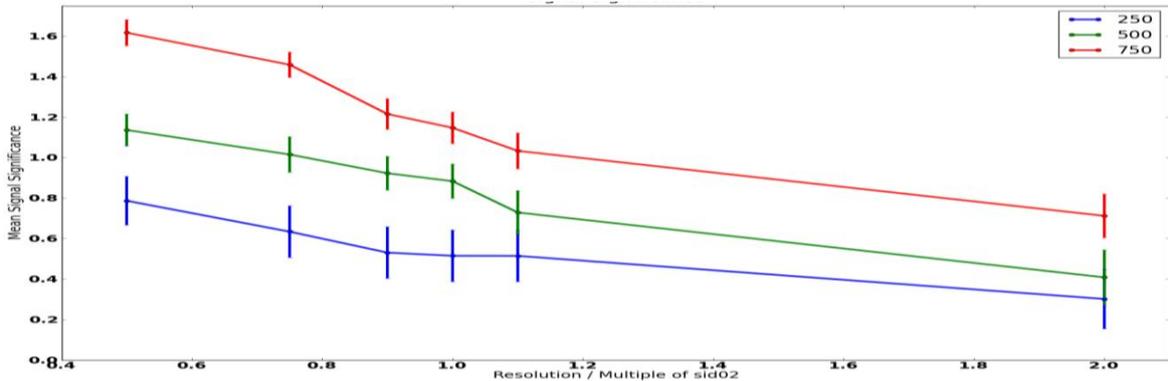

Figure 7. Mean significance level from an unbinned maximum likelihood fit in 250 fb$^{-1}$ (blue), 500 fb$^{-1}$ (green) and 750 fb$^{-1}$ (red) of data.

### 3.2. $Z^0 \rightarrow \mu^+\mu^-$ decay channel results

Like the "fast MC" results initially a proof of success is used to check if the segmentation alterations work as expected. In figure 8 hit distributions show the difference in position between the simulated hit in GEANT4 and the hit after the segmentation process. This has been shown for three different lengths of 25 μm wide strips. Figure 8(a) is the unchanged improvement in the transverse direction (due to the strip width remaining unchanged) and figure 8(b) is the improvement in the z direction due to the reduction in strip length. The width of these top-hat functions successfully represents the strip length.

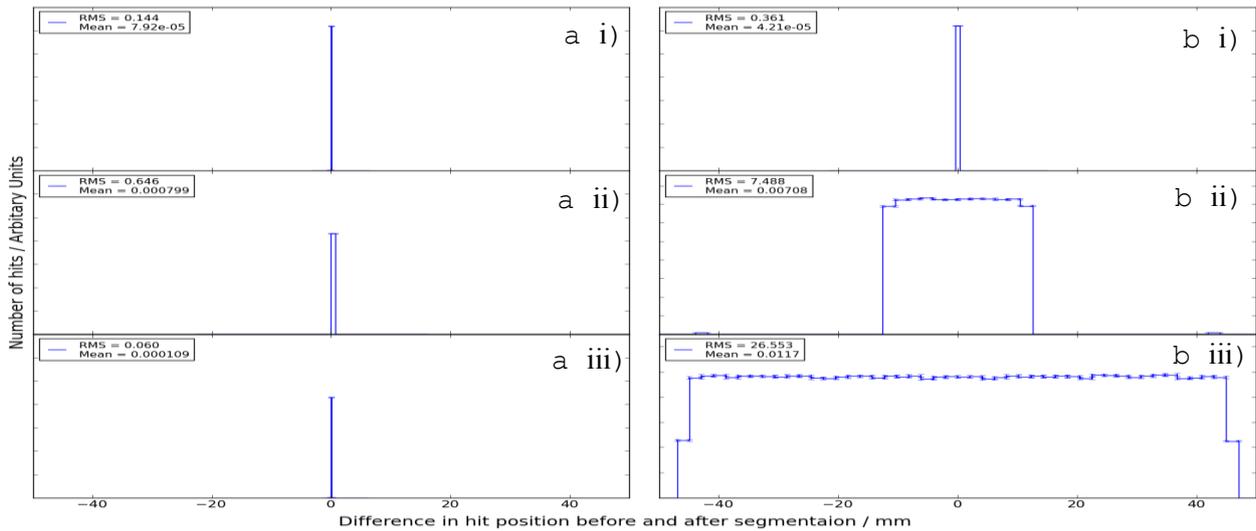

Figure 8. Hit distributions due to segmentation in the (a) x/y direction and (b) z direction for 25 $\mu m$ wide strips by 25 μm (i), 25 mm (ii) and 9.15 cm (iii).

Figure 9 shows the resolution of two of the track parameters. There is relatively no change in $\omega$ representing track curvature (See 2.3) and therefore little change in track momentum resolution; this was checked using a similar resolution graph for particle momentum. The resolution of the $z_0$ track parameter however is halved by the reduction in length of the strip from 9.15cm to 25um. This is expected as the $z_0$ track parameter is the distance in the z direction from an arbitrary reference point to the helix and the addition of pixels improves the z direction resolution in particular. Across all parameters however there is very little

Performance studies of a pixel tracker in the Silicon Detector (SiD) concept for a future linear collider change in the reconstructed tracks. The reconstruction of the $Z^0 \to \mu^+\mu^-$ events is improved very little by increased tracker granularity. Charged particles that pass through the vertex detector already have 5 hits that can be used for track reconstruction; any further hits in the tracker make no significant impact on resolution. As the vertex detector provides the limiting factor a greater understanding of the tracker performance should be achieved by using long lived particles that miss vertex layers.

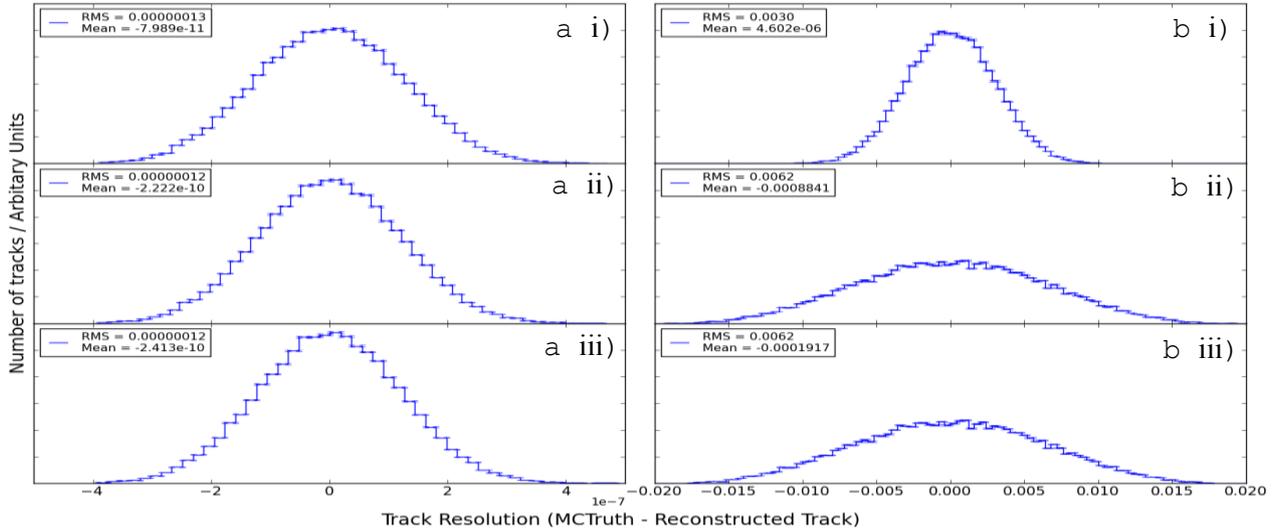

Figure 9. Track parameter resolutions for (a) $\omega$ and (b) $z_0$ for 25 $\mu m$ wide strips by 25 μm (i), 25 mm (ii) and 9.15 cm (iii).

## 3.3. $K_S^0 \to \pi^+\pi^-$ decay channel and stability studies

$K_S^0$ reconstruction is measured on its success by analysing the position of the secondary vertex. The resolution is the mean value of the distance between the MC and reconstructed vertex position across a large dataset. This resolution is plotted in figure 10 however it is important to note that at almost all lengths on this graph the strips outperform the pixels as they may have the same length but the pixels (same length and width) are generally wider creating a larger surface area. Looking at the entire detector (figure 10a) there is an improvement in resolution of approximately 2 orders of magnitude. However it is difficult to analyse values past 5000 μm where the track finding strategy begins to break down as the number of reconstructed tracks decreases, see figure 10b. As such it is only possible to conclude that any decrease in pixel size will only increase resolution down to about 100 μm. Below 100 μm there are other limiting factors that prevent further improvement, although this may be a feature of the simulation.

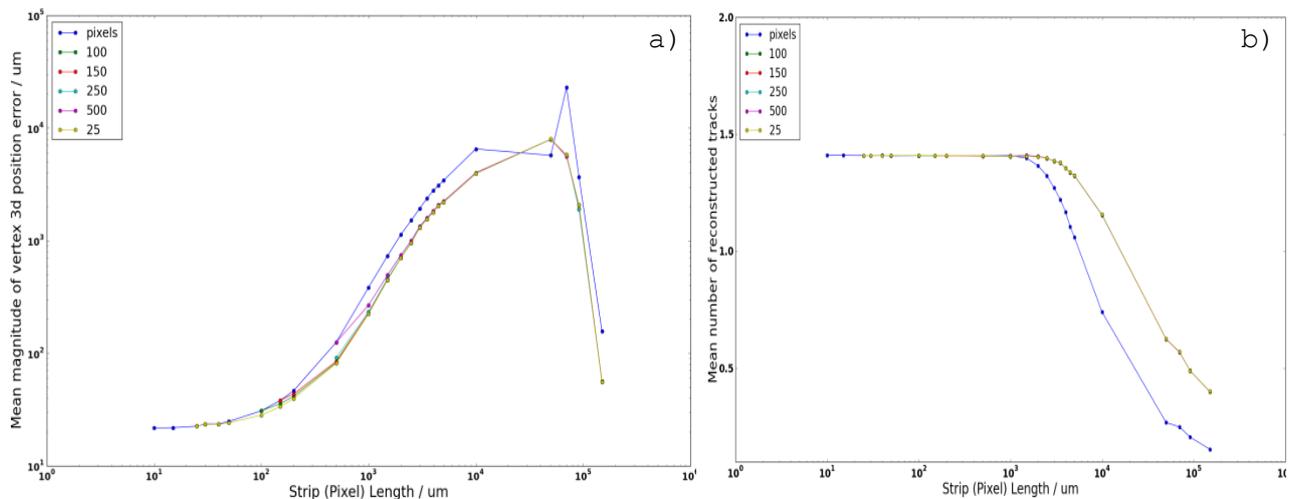

Figure 10. (a) Vertex displacement and (b) number of reconstructed tracks from the $K_S^0 \to \pi^+\pi^-$ decay channel across the entire detector.

A new reconstruction strategy specifically focussed on the forward region of the detector reconstructs more successfully. Some features also become clearer when focusing on a particular section of the detector, either the barrel or the endcap. Figure 11 shows the vertex resolution for the forward region and hence only

Performance studies of a pixel tracker in the Silicon Detector (SiD) concept for a future linear collider shows the endcap features. Firstly it's possible to see the strips 'joining' to the pixel line at the appropriate values (the strip length is reduced to a value equal to the width). The strip resolution line follows the pixel line closely until the length is about 4 times the width when the resolution improves abruptly. This is the point at which the current reconstruction software treats the z component as being 'bound' instead of an absolute value. This was detailed in 2.3. The final significant feature occurs around 9.15 cm. This is the original strip length and hence optimized with the geometric design. As the segmentation process 'tiles' the sensors onto the different layers if this is not an integer it is rounded down leaving insensitive material. This is the cause of this small but noticeable dip that becomes negligible when tiling micrometer sized pixels. Going from the current 9.15 cm x 25 μm strips to 100 μm pixels increases the secondary vertex resolution by a factor of 100 but also an increase in the number of channels by a factor of 226.

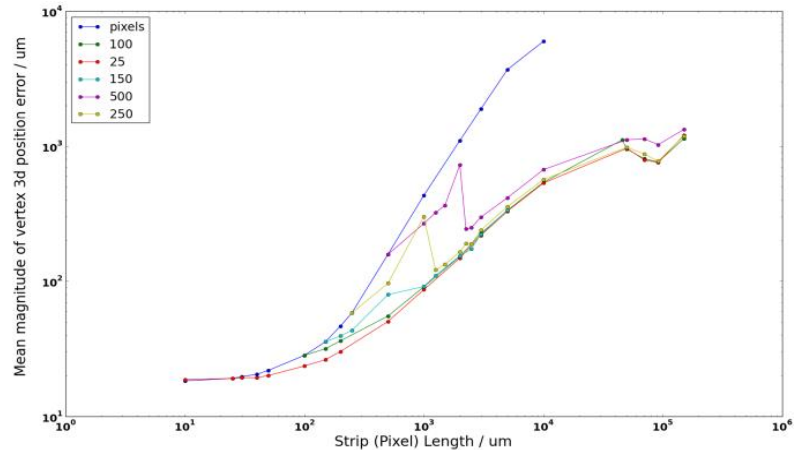

Figure 11. Vertex Displacement in reconstruction of the $K_S^0 \rightarrow \pi^+\pi^-$ decay channel in the forward region .

*3.4. Highly collimated dense jets*
Figure 12(a) shows that the number of hits in dense jet reconstruction decreases with increased length/width of the strips/pixels. This is expected as a sensor with a larger surface area is more likely to experience multiple hits. A noticeable feature is that the number of hits drops significantly for strips once the length is greater than 4 times the width. This is again a feature of the simulation, most probably the combining of two strip hits to create a single hit at the point where the two strips cross. However the pixel results show a steady decrease in number of hits with increased size. Figure 12(b) shows the number of tracks reconstructed and shows that the drop in number of hits for strips does not create a drop in number of tracks. This suggests that the reduction in number of hits is part of the reconstruction and not due to increased sensor dimensions. What is shown is the reduction in number of tracks for the pixels. 9.15 cm strips perform similarly to 10 μm pixels, hence strips are thin enough and strip reconstruction robust enough to allow successful reconstruction of the tracks.

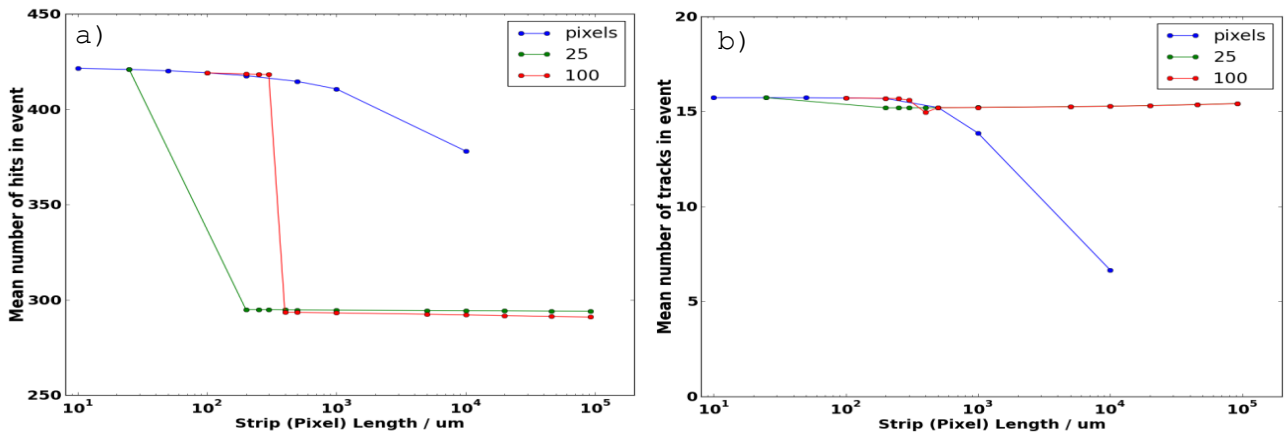

Figure 12. (a) Number of hits and (b) number of tracks in highly collimated dense jet events.

To analyse the quality of the tracks they will be matched with the corresponding MC particle using a simple truth matching strategy where 80% of the hits from the track must come from a single MC particle. The momentum difference calculated and the resolution (RMS value from a large number of events) is plotted in figure 13. There is an increase in resolution as the strips are reduced but this only becomes apparent once the length approached the width (pixels). The pixels also show an uncharacteristic improvement in resolution for larger sizes; this is probably due to the reduction in number of tracks reconstructed shown in figure 12b which invalidates the results. The resolution of the 25 µm wide strips doubles from 9.15 cm to 25 µm. This increase in momentum resolution would lead to higher signal significance for dense jet events as suggested by the $H \rightarrow \mu^+\mu^-$ study. The most common Higgs decay (table 1) is to b quarks and through quark confinement effects these produce dense jets. Hence increased dense jet resolution could lead to a shorter time before a Higgs discovery. However as with the $K_S^0$ reconstruction the necessary increase in channel number does not match the increase in resolution obtained. In order to obtain the increase in the obtainable physics the tracker would need 25 µm pixels representing an increase by 3660 in the number of sensor channels needed.

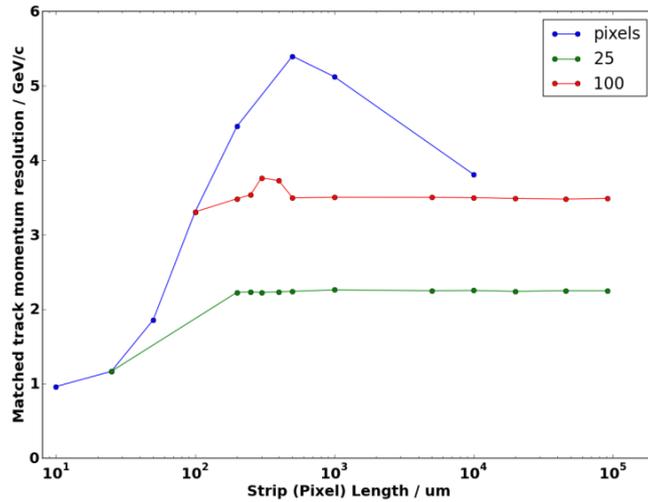

Figure 13. Root mean squared value of the momentum resolution for matched tracks (80% of track hits attributed to a single final state particle) in highly collimated dense jet events.

**Conclusions**

A "fast MC" based study was carried out on the $H \rightarrow \mu^+\mu^-$ decay channel and this showed considerable gain in signal significance could be achieved as a result of increased charged particle momentum resolution. As with previous studies upgrading the momentum resolution can lead to significantly improved physics results. However this improvement in resolution was shown to be unfeasible when a full simulation of the $Z \rightarrow \mu^+\mu^-$ decay channel was analysed. An increase in tracker granularity provided no significant improvement in momentum resolution, although an improvement was seen in the z component of the track parameters. Charged particles which passed through the vertex detector already contain 5 hits that can be used for reconstruction; any subsequent hits in the tracker make no significant impact.

The improved performance shown in the "fast MC" study is not possible with short lived charged particles. Conversely detector stability studies into missing/dead vertex layers using long lived particles showed greater improvement with increased tracker granularity. Using the $K_S^0 \rightarrow \pi^+\pi^-$ decay channel it was shown that there was no improvement in track reconstruction below about 100 µm pixels. Hence the existing 9.15 cm x 25 µm silicon strip geometry was replaced with 100 µm x 100 µm silicon pixels which provided an improvement in secondary vertex resolution of the order of 100. This improvement shows that the introduction of a pixel tracker considerably improves the robustness of the detector against failure of vertex layers and in the reconstruction of long lived particles. However this improvement comes with a necessary increase in the number of additional sensor channels by a factor of 226. Dense jet studies into highly collimated events showed the momentum resolution can be increased by a factor of 2. This increase in resolution could lead to higher signal significance for these events, but requires a reduction in pixel size to 25 µm representing an increase the number of sensor channels by a factor of 3660.

The linear collider is designed to make precision measurements on the physics discovered at the LHC. In order to achieve this high precision the Higgs decay channels that will be studied the most are decay into two leptons, which allows more precision than studying jet events. In conclusion an upgrade of the tracker granularity from the 9.15 cm strips to any micrometer sized pixels would requires a increase in number and complexity of sensor channels yet in the majority of linear collider physics cases provide only a relatively small improvement. However it should be noted that the addition of a pixel tracker does greatly improve the

Performance studies of a pixel tracker in the Silicon Detector (SiD) concept for a future linear collider robustness of the detector allowing high precision measurements to be made without the use of the vertex detector.

**Acknowledgements**
I would like to thank the project supervisor Dr Marcel Stanitzki for all his help and direction throughout the project and Dr Jan Strube for his invaluable guidance in particle physics as well as a multitude of programming languages.